\documentclass[preprint,aps,prl,superscriptaddress]{revtex4-1}

\usepackage{amsmath}    
\usepackage{graphicx}   
\usepackage{verbatim}   
\usepackage{color}      
\usepackage{hyperref}   
\usepackage{dcolumn}
\usepackage{amssymb}
\usepackage{bm}
\usepackage{epsfig}
\usepackage{epstopdf}
\usepackage{psfrag}
\usepackage[usenames]{xcolor}
\begin{document}

\title{A Micrometer-sized Heat Engine Operating Between Bacterial Reservoirs}
\author{Sudeesh Krishnamurthy}
\affiliation{Department of Physics, Indian Institute of Science, Bangalore - 560012, INDIA}
\author{Subho Ghosh}
\affiliation{Molecular Biophysics Unit, Indian Institute of Science, Bangalore - 560012, INDIA}
\author{Dipankar Chatterji}
\affiliation{Molecular Biophysics Unit, Indian Institute of Science, Bangalore - 560012, INDIA}
\author{Rajesh Ganapathy}
\affiliation{International Centre for Materials Science, Jawaharlal Nehru Centre for Advanced Scientific Research, Jakkur, Bangalore - 560064, INDIA}
\author{A. K. Sood}
\affiliation{Department of Physics, Indian Institute of Science, Bangalore - 560012, INDIA}
\affiliation{International Centre for Materials Science, Jawaharlal Nehru Centre for Advanced Scientific Research, Jakkur, Bangalore - 560064, INDIA}
\date{\today}

\begin{abstract}

\textbf{Artificial micro heat engines are prototypical models to explore and elucidate the mechanisms of energy transduction in a regime that is dominated by fluctuations \cite{StirlingEffort, Hangii review}. Micro heat engines realized hitherto mimicked their macroscopic counterparts and operated between reservoirs that were effectively \textit{thermal} \cite{Bechinger Nature, Raul Rica, Szilard engine, Lutz Optical engine, Lutz Ion engine}. For such reservoirs, temperature is a well-defined state variable and stochastic thermodynamics provides a precise framework for quantifying engine performance \cite{Sekimoto, Seifert review}. It remains unclear whether these concepts readily carry over to situations where the reservoirs are out-of-equilibrium \cite{Sriram transport}, a scenario of particular importance to the functioning of synthetic \cite{Browne Synthetic Molecular Motors, Balzani} and biological \cite{Howardj} micro engines and motors. Here we experimentally realized a micrometer-sized \textit{active} Stirling engine by periodically cycling a colloidal particle in a time-varying harmonic optical potential across bacterial baths at different activities. Unlike in equilibrium thermal reservoirs, the displacement statistics of the trapped particle becomes increasingly non-Gaussian with activity. We show that as much as $\approx$ 85\% of the total power output and $\approx$ 50\% of the overall efficiency stems from large non-Gaussian particle displacements alone. Most remarkably, at the highest activities investigated, the efficiency of our quasi-static active heat engines surpasses the equilibrium saturation limit of Stirling efficiency - the maximum efficiency of a Stirling engine with the ratio of cold and hot reservoir temperatures ${T_C\over T_H} \to 0$. Crucially, the failure of effective temperature descriptions \cite{Bocquet, Cates,Roberto Leonardo} for active reservoirs highlights the dire need for theories that can better capture the physics of micro motors and heat engines that operate in strongly non-thermal environments.}
\end{abstract}

\maketitle

In a seminal experiment, Blickle and Bechinger \cite{Bechinger Nature} devised a micrometer-sized Stirling heat engine that was driven by fluctuations from an equilibrium thermal reservoir, while being subject to time-dependent potentials in an optical trap. Like its macroscopic counterpart, the mean quasistatic efficiency of such an engine ($\approx$ 14$\%$ for a colloidal bead in water) is given by $\mathcal{E}_\infty = \mathcal{E}_c\left[1+\mathcal{E}_c / \text{ln} \left[\frac{k_{max}}{k_{min}}\right]\right]^{-1}$. Here, $\mathcal{E}_c = 1 - {T_{C}\over T_{H}}$ is the Carnot efficiency, $T_C$ and $T_H$ are the cold and hot reservoir temperatures, respectively, and the ratio of trap stiffnesses $k_{max}\over k_{min}$ is analogous to the compression ratio of a macroscopic engine. By imposing an external source of noise on  the trapped colloidal particle, corresponding to an effective $T_{H}$ of nearly 3000 K \cite{Petrov}, and implementing the microscopic equivalent of an adiabatic process, a Brownian Carnot engine with higher power output and efficiency was later realized \cite{Raul Rica}. Although strategies for harnessing the work done by these micro heat engines have not yet been devised, these studies underscore the feasibility of using a colloidal particle as the working substance of a heat engine to elucidate the role of fluctuations on its performance. A common feature of micro heat engines investigated hitherto, is that the noise fluctuations associated with the reservoirs decorrelate on a time scale that is much smaller than that associated with fluctuations of the working substance - the Brownian time of the colloidal bead. Owing to this separation of time scales, particle dynamics in such reservoirs follow Gaussian statistics and equilibrium stochastic thermodynamics can be readily applied. However, in active reservoirs - a bath of self-propelled particles - a canonical out-of-equilibrium system, noise fluctuations remain temporally correlated for substantially longer. The displacements of the trapped colloidal particle thus follow non-Gaussian statistics over suitable time and length scales \cite{Cates,Roberto Leonardo}. Whether incorporating effective temperature descriptions of active noise \cite{Bocquet} within stochastic thermodynamics can successfully describe the functioning of these engines remains unclear.   

Here we designed, constructed and quantified the working of an \textit{active} heat engine.  A 5$\mu$m colloidal bead, held in a harmonic optical trap, in a suspension of motile bacteria - \textit{B. licheniformis} - acts as the working substance. The time-dependent variations in laser intensity, i.e. trap strength, mimic the role of the piston of a macroscopic engine. A key advantage of utilizing a bacterial reservoir is that bacterial metabolism and the corresponding activity are strongly sensitive to the bath temperature \cite{Shneider,Lewis}. We exploit this behaviour and periodically create conditions of high and low activity, thus imitating passive reservoirs with a temperature difference (see Materials and Methods and supplementary text and Figs. S1-S2). The bacterial activity depends on a host of physico-chemical parameters and was found to be exclusive to each experiment. This allowed us to access reservoirs with different activities, while keeping the bacterial number density constant. The engine is driven by fluctuations arising from bacterial activity and we execute the microscopic equivalent of a Stirling cycle.
 
\begin{figure}[tbp]
\centering
\includegraphics[scale=0.42]{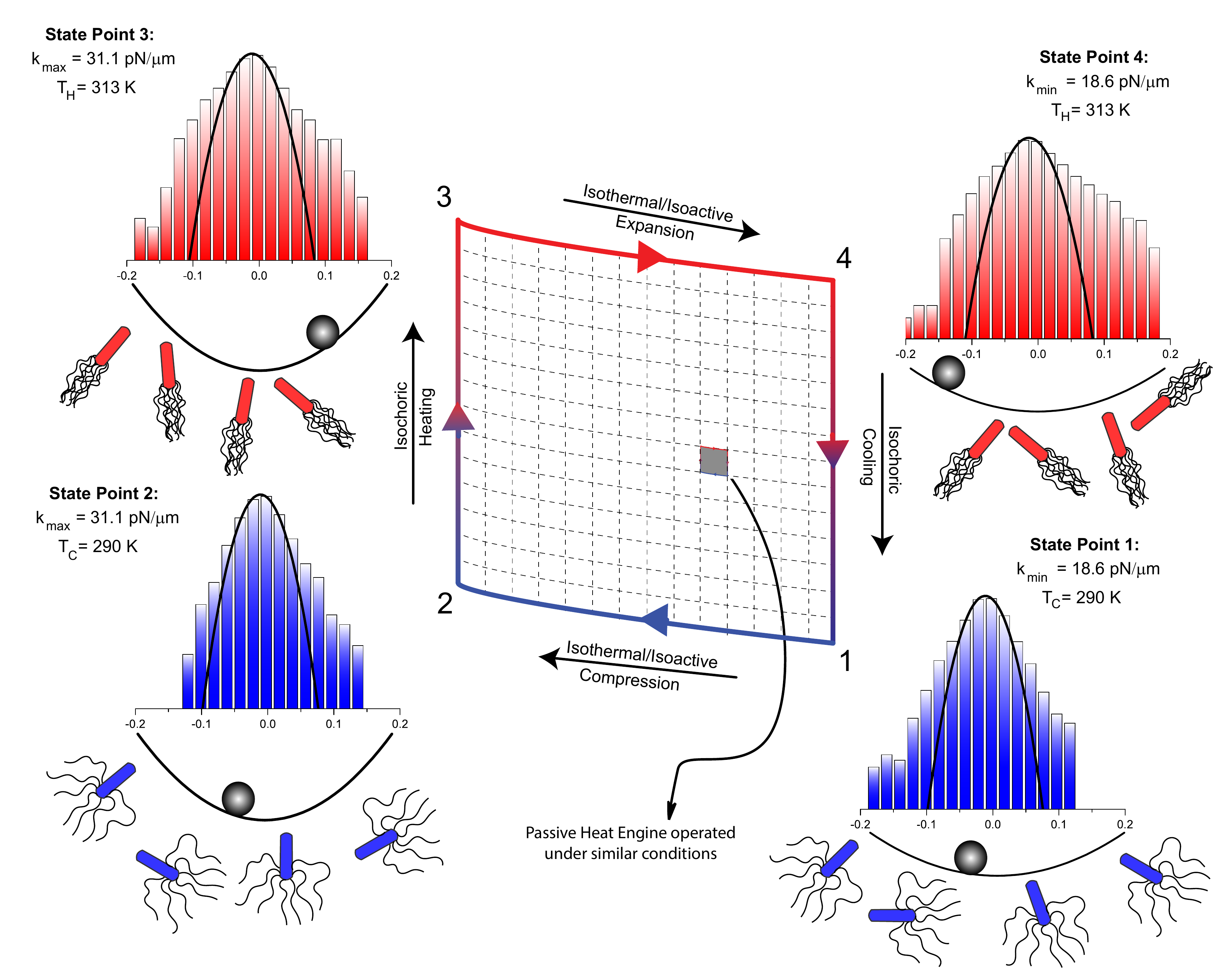}
\caption{\textbf{Micrometer active Stirling engine}. A Stirling cycle consists of isothermal compression ($1\rightarrow2$) and expansion ($3\rightarrow4$) steps at temperatures $T_C$ and $T_H$, respectively, connected by isochores $2\rightarrow3$ and $4\rightarrow1$. In the active Stirling engine, for the highest activities investigated, a substantial contribution to the total work stems from bacterial activity and the isotherms can be effectively replaced by isoactivity lines. Increasing (decreasing) the trap stiffness decreases (increases) the phase space volume available to the colloidal particle and mimics a compression (expansion) stroke of a macroscopic Stirling engine. $P(\Delta x)$ at state points 3 and 4 are substantially more non-Gaussian that state points 1 and 2 due to increased bacterial activity. The black lines represent Gaussian fits. The area of the large(small) Stirling cycle represents the average work done by the active(passive) engine as it executes one Stirling cycle. }
\label {Figure1}
\end{figure}

A typical cycle executed by the engine is outlined in Fig. \ref{Figure1}. The process starts at state point 1 where $T_{C} = 290$ K and trap stiffness, $k = k_{min} = 18.6$ pN/$\mu$m (see supplementary Fig. S3). Bacteria in this state are sluggish but still active. The probability distribution of displacements in the x-position, $P(\Delta x)$, of the colloidal bead corresponding to this state point is also shown. The black line is a Gaussian fit to the data and points beyond the fit represent non-Gaussian displacements. $k$ is increased linearly to $k_{max} = 31.1$ pN/$\mu$m analogous to a macroscopic isothermal compression and the system reaches the state point 2. The increase in $k$ results in a narrower $P(\Delta x)$. The bath temperature is increased to $T_{H} = 313$ K at $k = k_{max}$ and the system reaches state point 3. A substantial increase in the width of $P(\Delta x)$ due to enhanced bacterial activity during the isochoric heating is clearly evident (Figure \ref{Figure1}).  $k$ is decreased linearly to $k_{min}$ analogous to a macroscopic isothermal expansion and the system reaches state point 4. Finally, the cycle returns back to state point 1 by isochoric cooling to $T_{C}$ keeping $k = k_{min}$. $T_{C}$ and $T_{H}$ are kept the same for the passive (no bacteria present) as well as the active engine for all activities investigated. The isothermal processes were executed in 7 s and isochoric ones in 4 s and the time taken to complete one Stirling cycle, $\tau = 22$ s was held constant (see Supplementary Information). The position of the colloidal bead represents the state of the system as the cyclic process described in Fig. \ref{Figure1} is steadily executed.

Thermodynamic quantities were computed from particle trajectories using the framework of Stochastic Thermodynamics \cite{Seifert review,Bechinger First law}. The work done by the engine is $W = \int \frac{\partial U}{\partial t} dt$ where $U(x,t) = {1 \over 2}k(t)x(t)^2$ with $x(t)$ being the displacement of the bead from the trap center at time $t$. In this sign convention, work done,$W$ by(on) the system on(by) the surrounding is negative(positive). $W$ is non-zero during the isothermal processes (1 $ \rightarrow $ 2 and 3 $ \rightarrow $ 4)and zero in the isochoric processes( 2 $ \rightarrow $ 3 and 4 $ \rightarrow $1). Heat transferred, $Q$ by the reservoir in the isothermal processes can be calculated using an energy balance reminiscent of the First law of thermodynamics $dU = W - Q$ , despite $Q$ and $W$ being fluctuating quantities \cite{Bechinger First law}. The energy balance is observed to be true for our active engines as well (See discussion following Eq 6 of Supplementary Information). In the isothermal processes $dU =0$ and $Q = W$. Heat transferred during the isochoric processes is calculated using $Q =-\int \frac{\partial U}{\partial x}\dot{x}dt$, and for constant $k$, results in a path independent form for $Q = -\frac{k}{2}\left[ x^2\right] ^t_0$. The efficiency of the engine defined as, $\mathcal{E} = W_{cycle}/Q_h$, where $Q_h$ is the heat transferred in the isothermal expansion 3 $ \rightarrow $ 4 and $W_{cycle}$ is the work done at the end of the Stirling cycle.

\begin{figure}[tbp]
\centering
\includegraphics[scale = 0.6,trim={4cm 4cm 4cm 0},clip]{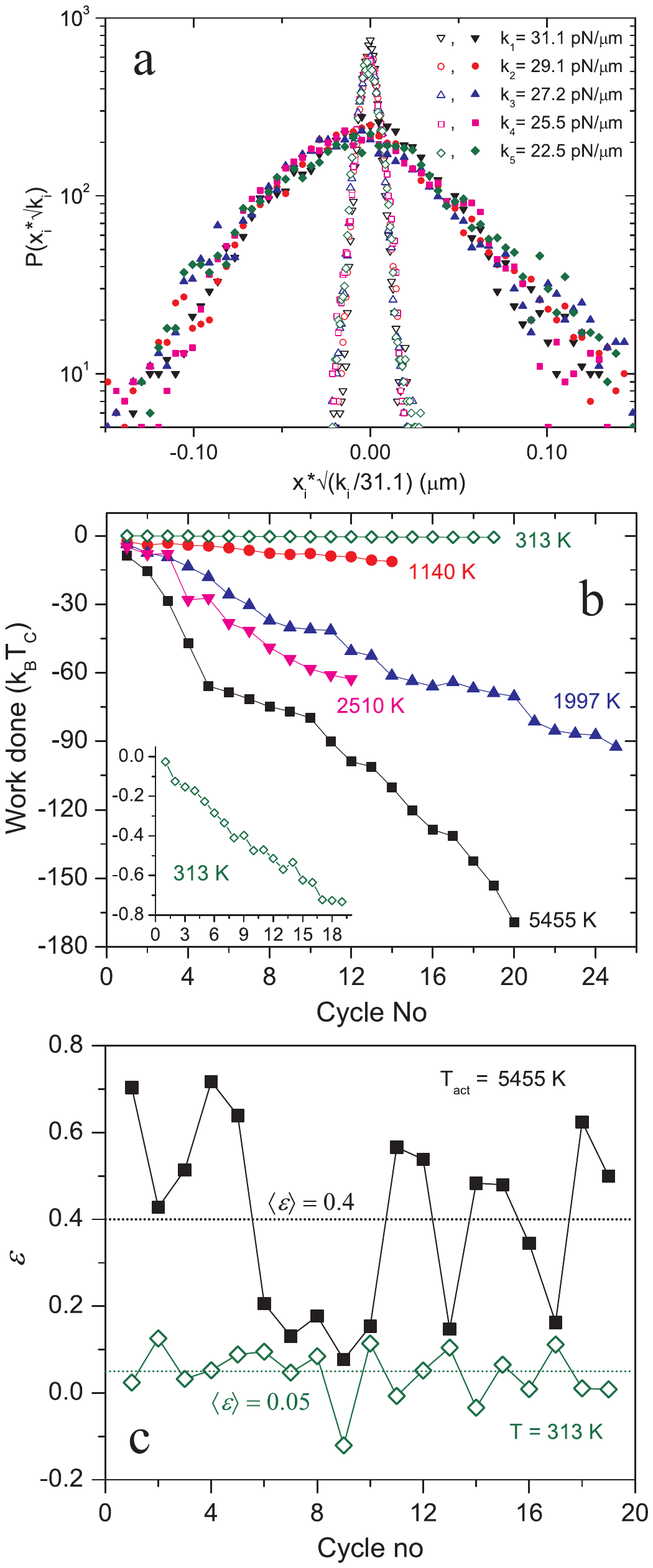}
\caption{\textbf{Comparison of active engines at different activities.}
\textbf{a} Scaled $P(\Delta x)$'s at different instances along the isothermal expansion stroke (3$\rightarrow$4) for two different activities. $T_{act} = 5455$ K (closed symbols) and $T_{act}= 1140$ K (open symbols) \textbf{b} Cumulative work done by active (closed symbols) and passive (open symbols) engine. The temperatures denote $T_{act}$ of the hot reservoir. Inset to \textbf{b} shows work done by the passive engine on a scale where fluctuations are visible. \textbf{c} Efficiency of the active $T_{act} = 5455$ K-(closed symbols) and passive (open symbols) engine as the Stirling cycle is continuously executed. The dotted black (green) line represents the mean efficiency of the active(passive) engine.}
\label{Figure2}
\end{figure}

We first compared the performance of an active engine with a passive one (see supplementary movies SM1 and SM2). For the highest activity accessible in our experiments, the power $\mathcal{P} = W/\tau$, of the active engine, represented by the area of the larger Stirling cycle in Fig. \ref{Figure1}, is over two orders of magnitude larger than that of the passive engine (see supplementary Figs. S4 and S5). $\mathcal{P}$ for a heat engine is a non-monotonic function of $\tau$. As $\tau\to0$, a substantial amount of heat drawn from the hot reservoir is lost towards irreversible work, $W_{irr}$, and $\mathcal{P}$ is small since $W$ is small. For large cycle durations ($\tau\to\infty$), i.e. the quasistatic limit, while both $W_{irr}$ and $\mathcal{P}\to0$, $\mathcal{E}$ reaches its maximum as the heat drawn during the isotherms is completely converted to work. Performance of engines in this limit is described by equilibrium stochastic thermodynamics. For intermediate $\tau$, however, a trade-off between $W_{irr}$ and $\tau$ results in a maximum in $\mathcal{P}$ \cite{Seifert model,Curzon Ahlborn}. Thus, a comparison across engines is only possible at maximum $\mathcal{P}$ or in the quasistatic limit (maximum $\mathcal{E}$), where the nature of irreversibility is well-understood. In the quasistatic limit the instantaneous $P(\Delta x)$ of the colloidal particle should mimic the closing (1 $\rightarrow$ 2) and opening (3 $ \rightarrow $ 4) of the optical trap. This implies, $k(t_1)\langle\Delta x^2(t_1)\rangle = k(t_2)\langle\Delta x^2(t_2)\rangle = k_B T$ where, $t_1$ and $t_2$ denote any two time instances along the isotherms. Thus the $P(\Delta x)$'s (determined over suitably small time bins), when appropriately scaled should collapse and this is indeed observed for $\tau = $ 22s (Fig. \ref{Figure2}a) (see also Supplementary Fig. S6 ). The engines are thus operating in the quasistatic limit, which enables a direct comparison of engine performance across activities.    

In Fig. \ref{Figure2}b, we show the total work done, $W_{total}$ in units of $k_B T_C (T_C = 290K)$, at the end of each Stirling cycle by the passive engine (hollow symbols) and the active engine, for various bacterial activities (solid symbols). The slopes of the trajectories are negative and work is done by the engine on the surroundings. To represent various bacterial activities, we define an active temperature $k_{B}T_{act} = \frac{1}{2}k_x \langle x^2\rangle +\frac{1}{2}k_y \langle y^2\rangle  = \langle U\rangle$. Fluctuations in our system follow non-Gaussian statistics and $T_{act}$ is not the effective temperature of the bacterial bath. Rather, we use $T_{act}$ only to qualitatively parametrize different activities. As per the definition, $T_{act}$ is the temperature of an equilibrium reservoir with the same average potential energy $ \langle U\rangle$ as the trapped bead in our bacterial reservoir. As a consequence, the equilibrium reservoir at $T_{act}$ and our bacterial reservoir transfer the same amount of heat during an isochoric process. With increasing $T_{act}$, $W_{cycle}$ steadily increases (Fig. \ref{Figure2}b). At the highest activity investigated, the hot and cold reservoir $T_{act}$'s are $\approx 5500$ K and $\approx 2300$ K, respectively. Such enormous differences in reservoir temperatures are impossible to mimic in passive microengines without an external source of noise. $W_{cycle}$ is a fluctuating quantity due to the stochastic nature of the engine and so is the efficiency, $\mathcal{E}$. Figure \ref{Figure2}c shows these fluctuations for a engine in contact with a passive (green diamonds) and the active(black squares) reservoir. Most remarkably, at the highest activity, $\langle\mathcal{E}\rangle$ of the bacterial engine is almost an order of magnitude larger than its passive counterpart and is only a factor of two smaller than biological motors.    

\begin{figure}[tbp]
\centering
\includegraphics[scale = 0.75,trim={0cm 10cm 1cm 2cm},clip]{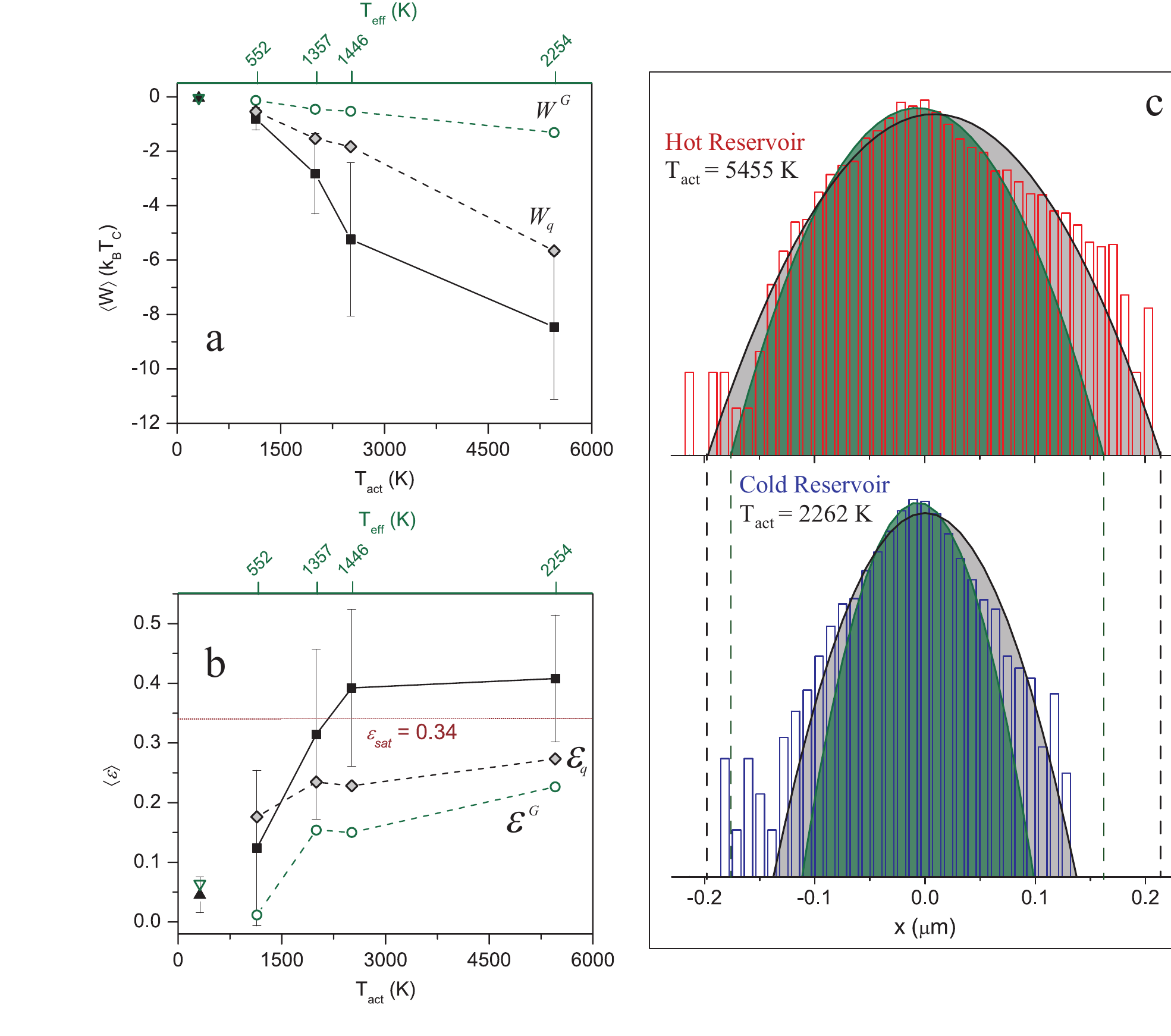}
\caption{\textbf{Elucidating the origins of active engine performance.}  \textbf{a} and \textbf{b} show the work done and efficiency for the experimental and simulated engines, respectively. Experimental and calculated $W_{cycle}$ and $\mathcal{E}$ for passive engines are represented by solid triangles and hollow inverted triangles respectively.  Solid squares represent experimental active engine, hollow circles represent simulated Gaussian engine and diamonds represents the quasistatic Stirling engine. The efficiency of passive engines saturates for $T_C/T_H \rightarrow 0$ at $\mathcal{E}_{sat}$ represented by the brown line in \textbf{b}. \textbf{c}  Experimental $P(\Delta x)$'s at trap stiffness, $k = k_{min}$, in the hot (red bars) and cold (blue bars) reservoirs for the highest activity studied. The green line represents the Gaussian fit to the data and the fluctuations in the green shaded region alone contribute to $W^G$ and $\mathcal{E}^G$ (green circles in \textbf{a} and \textbf{b}). The black line represents a Gaussian engine with the same variance as the experimental active engine. Fluctuations within the grey shaded area contribute to $W_{q}$ and $\mathcal{E}_q$. }
\label {Figure3}
\end{figure}

The $\langle W_{cycle}\rangle$ and $\langle\mathcal{E}\rangle$ for the passive (solid triangle) and the active engine at different $T_{act}$'s (solid squares) are shown in Figs. \ref{Figure3}a and \ref{Figure3}b, respectively. In the quasistatic limit, the work done per cycle is given by $W_q = 2 k_BT_c\left[1-\frac{T_H}{T_C}\right] \text{ln} \sqrt{\frac{k_{max}}{k_{min}}}$. $W_q$ and $\mathcal{E}_q$ for the passive engine, represented by hollow inverted triangles in Fig. \ref{Figure3} a and b, closely match the experimental values (solid triangles) and reaffirm that we are operating the engine in the quasistatic limit. Turning our attention to the active engines, we find that while $\langle W_{cycle}\rangle$ increases monotonically with $T_{act}$ (Fig. \ref{Figure3}a), $\langle\mathcal{E}\rangle$ increases and then saturates at high $T_{act}$'s (Fig. \ref{Figure3}b). Such a behaviour is typical of an equilibrium Stirling engine where the quasistatic efficiency saturates to $\mathcal{E}_{sat} = (1 + 1/ \text{ln}(k_{max}/k_{min}))^{-1}$ as $T_C/T_H \to 0$. $\mathcal{E}_{sat}$ for our experiment is represented by the dashed horizontal line in Fig. \ref{Figure3}b. The experimentally determined efficiencies surpasses this limit suggesting a failure of the equilibrium description in evaluating $\mathcal{E}_{sat}$ for the active engines.

To understand the origin of such high $\mathcal{P}$ and $\mathcal{E}$ of our active engines, we carefully examined the fluctuations of the trapped colloidal particle in the active reservoirs. $P(\Delta x)$ in the hot(cold) reservoir at the highest activity studied is shown by the red(blue) histogram in Fig. \ref{Figure3}c. The green line represents a Gaussian fit to $P(\Delta x)$. In order to isolate the contributions from Gaussian fluctuations to the total $W_{cycle}$ and $\mathcal{E}$, we simulated Stirling cycles by drawing particle positions at random from the Gaussian region alone (green shaded area). For such a Gaussian engine, an effective temperature $T_{eff}$ can be precisely defined and $T_{eff} < T_{act}$ (top x-axis in Figs. \ref{Figure3}a and b) since we have chosen only the low energy contributions to the total $\langle U\rangle$. $\langle W^{G}\rangle$ and $\langle \mathcal{E}^{G}\rangle$ for the simulated Gaussian engine are shown as green circles is Figs. \ref{Figure3}a and b, respectively. Strikingly, at the highest activity, 85$\%$ of total $\langle W_{cycle}\rangle$ and 50 $\%$ of the total $\mathcal{E}$ is due to non-Gaussian fluctuations. In fact, these fluctuations account for 80$\%$ of the potential energy difference between the hot and cold reservoirs (Fig\ref{Figure3}c) explaining their large contribution to $\langle W_{cycle}\rangle$.

We can further compare the experimental active engine with a passive engine operating between equilibrium reservoirs at the same $\langle U\rangle$, i.e. the same $T_{act}$. The solid black curve in Fig \ref{Figure3}c represents a Gaussian with width equal to the variance of $P(\Delta x)$ of the active engine. The black diamonds in Figs. \ref{Figure3} a \& b, show $W_{q}$ and $\mathcal{E}_{q}$, respectively, for a quasistatic equilibrium Stirling engine operating under such conditions. $W_{q}$ and $\mathcal{E}_{q}$ continue to remain smaller than $\langle W_{cycle}\rangle$ and $\mathcal{E}$ of the corresponding active engine. Most strikingly, at the highest $T_{act}$ studied, 30$\%$ of the total work done is accounted for by a few large displacements seen outside of the grey shaded area of the Fig \ref{Figure3}c. These fluctuations comprise about 22$\%$ of the difference in $U$ between the $P(\Delta x)$ of the hot and cold reservoirs. Thus, despite having the same $\langle U\rangle$, owing to a few large non-Gaussian displacements that occur with a very low probability, active engines fare better than equilibrium ones.      

The non-Gaussian fluctuations of particles in active reservoirs is a result of the underlying bacterial motility.  It was recently shown that a colloidal particle diffusing in an active bath in the absence of any confining potential shows super-diffusive behavior for times smaller than a characteristic time, $\tau_c$ \cite{Wu Libchaber,Sriram superdiffusion}. In our experiment, $\tau_c \sim1s$ (see supplementary information and Fig. S7), while the time scale over which the particle experiences the optical trap is of the order of 10 ms (see supplementary materials). Further, exact solutions \cite{Cates} to the Langevin equation for a self propelled particle in a harmonic trap suggests that the shape of $P(\Delta x)$ is determined by parameter $a = (2\mu k \tau_c)^{-1}$, where $\mu$ is the mobility of the particle. $P(\Delta x)$ is strongly non-Gaussian for $a\ll1$ and in our experiment, $a\sim10^{-4}$.

Collectively, our results show that active engines significantly out perform passive engines, which are bound by the laws of thermodynamics. While stochastic thermodynamics provides a firm basis to interpret the performance of engines operating between equilibrium thermal reservoirs, we find that a naive application of effective temperature arguments\cite{Bocquet} to quantify activity, and the subsequent application of stochastic thermodynamics fails to capture the physics of active engines. Our experiments establish that a major contribution to the superior performance of active heat engines arises from non-Gaussian fluctuations\cite{Cates, Roberto Leonardo} in position and hence velocity (i.e. departure from the equipartition theorem) and these should be explicitly taken into account in future theories. Further, given the lack of measures for parameterizing active reservoirs, the observed sensitivity of active engine performance on the statistics of rare events suggests that they can be used as precise probes of activity. Although biological motors are known to typically operate under isothermal conditions\cite{Howardj}, it is tempting to speculate that gradients/ synchronization of activity within the cell also contributes to their observed high performance. In light of the recent advances in the fabrication of micro and nanoparticles that can be rendered active\cite{Aranson} by chemical\cite{Golestanian}, optical\cite{Sano}, electric and magentic fields\cite{Bibette}, active reservoirs can readily provide a self-contained source of noise. Therefore, such reservoirs should be an integral part of the design of future microscopic heat engines that can potentially power micro and nano electro-mechanical devices. 

\section*{Materials and Methods}
\subsection{Bacterial strain and growth condition}
\textit{B. licheniformis} (Weigmann 1898) Chester 1901 (also known as Bacillus licheniformis ATCC 14580) was bought from Microbial Culture Collection at National Centre for Cell Science, Pune, India (Catalog No.-MCC 2047). Bacteria were grown in Tartoff-Hobbs Hiveg broth (Himedia) at 37°C with vigorous shaking. Suspension of bacteria was aliquoted for experiment at sixth hour of growth in 5 ml of Tartoff-Hobbs Hiveg broth. 

\subsection{Laser Trapping and Particle position determination}
The crosslinked Poly(styrene/divinylbenzene) (P[S/DVB]) particles of 5 $\mu$m were obtained from Bangslabs, USA. The particles were trapped in a optical trap obtained by tightly focusing an IR laser beam with a 100x objective. A NdYVO4 laser of wavelength 1064 nm was used to produce the trapping laser beam. The laser beam was focused using a 100X Carl Zeiss objective (1.4 N.A.) mounted on a Carl Zeiss Axiovert Microscope. An extremely low power aligned red laser (Thorlabs ML101J8 Diode laser of wavelength 632nm controlled using a Thorlabs TCLDM9 temperature controlled laser diode module) is switched on during the isochoric process and is used as a marker to define the end points of the process with an accuracy within the limits of our temporal resolution (see Supplementary movie, SM1 and SM2). $T_{bath}$ is tuned by flowing a heat exchanging fluid, in this case water, in a channel adjacent to the bacterial reservoir. The bacterial reservoir and the heat exchanging fluid are separated by a \#1 (100 $\mu$m thick) glass coverslip for quick equilibration. Particles were imaged using a Basler Ace 180 kc color camera at 500 frames/sec. Only the green slice of the RGB image was considered to eliminate the influence of the red laser(632nm) on the particle position. The particle was tracked to sub-pixel resolution using the tracking codes by R. Parthasarathy \cite{Parthasarathy}. The particle positions could be determined with an accuracy of 10nm.

\section*{Acknowledgements}
We thank Udo Seifert for illuminating discussions and Sriram Ramaswamy and Shreyas Gokhale for critical comments on our manuscipt. AKS thanks J C Bose Fellowship of the Department of Science and Technlogy, India for support. SK thanks DST for fellowship support. RG thanks the ICMS, JNCASR for financial support.

\newpage

\newpage

\newpage

\end{document}